\documentclass{PoS}

\title{The effect of fluctuations of the vortex core size on the static
potentials evaluated by the thick center vortex model}

\ShortTitle{Fluctuation of the vortex core size}

\author{\speaker{Sedigheh Deldar}\\
        Department of Physics, University of Tehran, P.O. Box
14395/547, Tehran 1439955961, Iran\\
        E-mail: \email{sdeldar@ut.ac.ir}}

\author{Shahnoosh Rafibakhsh\\
        Department of Physics, University of Tehran, P.O. Box
14395/547, Tehran 1439955961, Iran\\
        E-mail: \email{rafibakhsh@ut.ac.ir}}

\abstract{By varying the vortex core size of the thick center vortex model, we have
studied the short distance potentials between static sources. It has
been found that fluctuations of the vortex core size lead to Coulombic
behavior. Furthermore, we discuss the influence of such fluctuations on
Casimir scaling for both the Coulombic and the linear part of the potential.\
        }

\FullConference{8th Conference Quark Confinement and the Hadron Spectrum \\
		 September 1-6 2008\\
		 Mainz, Germany}

\begin{document}

\section{Introduction}

The behavior of the potential between static sources helps to understand the nature of the forces in Quantum Chromodynamic. Thick center vortex model \cite{Faber:1998} is one of the phenomenological models that describes the potential at medium and large distances, fairly in agreement with the lattice data. The interaction between vortices and the Wilson loop leads to a linear potential proportional to Casimir scaling at intermediate distances, and to an asymptotic N-ality dependence at large distances. Based on the model, the Wilson loop and the induced potential between the heavy sources of the SU(N) gauge group have been obtained: 
\begin{equation}
<W(C)>= \exp[-\sigma(C)A]<W_{0}(C)> 
\label{W}
\end{equation}
where $W(C)$ is the Wilson loop for two quarks which are located at a distance $R$, $\sigma$ is the string tension and $A$ is the area of the loop. The exponential leads to a linear potential and it is confirmed by lattice calculations that the non-trivial center elements are responsible for this event \cite{Faber:1999}. The non-trivial center elements of each SU(N) gauge group are:
\begin{equation}
z_{n}=\exp({\frac{2\pi i n}{N}}) \qquad n=1,2, \cdots, N-1
\end{equation}
Making the vortices thick \cite{Faber:1998}, one gets the intermediate linear potential not only for the fundamental but also for higher representation sources. In addition, the large distance potentials have been obtained in agreement with the fact that the string tensions of zero N-ality representations must be zero and the representations with the same non-zero N-ality must acquire the same string tensions. These results are very impressive for a very simple model especially the agreement with Casimir scaling . 
The successes of the model have encouraged us to go one step further and to think about getting the correct short distance potential  from the model. If the interaction of the vortices with the Wilson loop gives the intermediate and large distance potentials, where the vortices are some special class of configurations that somehow explain the role of the gluonic field of QCD, why one may not consider some other special classes of them to be responsible for the short distance behavior where the Coulombic potential is obtained by one gluon exchange between the sources.

In the thick center vortex model, non-trivial center elements give the $\exp[-\sigma(C)A]$ of equation (\ref{W})  but there has been no work within the model to introduce a mechanism for reproducing $W_{0}$. One may introduce an operator which interacts with the Wilson loop and obtains the Coulombic part. This is what M. Faber and D. Neudecker  are doing and their primary results are presented in the poster session of this conference. We have chosen another approach and kept the center vortices as the operators, even for the short distances, but with some modifications. Vortices are known as the non-trivial center elements of the gauge group. We have tried to add back $W_{0}$ to $W(C)$ by increasing the role of the trivial center element which has been entered somehow in the model when the thick center vortices have been introduced. 

A brief review of the thick center vortex is presented in the next section. Then, in section three, we discuss the fluctuations of the vortex core size and its effect on the short distance potential. It is shown that with this method, one can get a qualitative agreement between the potentials and  Casimir scaling for all regimes, including the Coulombic part.

\section{Static sources potentials by thick center vortex mechanism}

The thick center vortex model \cite{Faber:1998} is a modification of the vortex model \cite{Hooft:1979}, which gives both the intermediate and large distances potentials for sources of the fundamental and higher dimensional representations. Based on the original thin vortex model, the non-trivial center elements of the corresponding gauge group were responsible for the linear potential at large distances. Therefore, the interaction of a vortex and a Wilson loop could be obtained by inserting one of the non-trivial center elements in the Wilson loop as the following:
\begin{equation}
W(C)=Tr[UU...U]\rightarrow Tr[UU...z...U].
\label{WU}
\end{equation}
where $z=\exp[{\frac{2\pi i n }{N}}]$. If there was no interaction between the vortex and the Wilson loop, one could insert a unit matrix of the gauge group, $I$, instead. This method gave the pattern of the large distance potentials correctly; but not the intermediate potentials especially for higher representations where the lattice calculations have predicted a linear potential. By thickening the vortices, this problem has been solved. The thin vortex  or the $z$ in equation (\ref{WU}) is replaced by a group element $G$ which interpolates smoothly from one center element to another one. It is equal to $+I$ if the vortex is not linked to the Wilson loop at all. The result of this modification is \cite{Faber:1998}:
\begin{equation}
<W(C)> = \prod_{x} \{ 1 - \sum^{N-1}_{n=1} f_{n} (1 - Re {\cal G}_{r}
          [\vec{\alpha}^n_{C}(x)])\}, \qquad {\cal G}_{r}[\vec{\alpha}] = \frac{1}{d_{r}} Tr \exp[i\vec{\alpha} . \vec{H}].
\label{WC}
\end{equation}
$x$ is the location of the center of the vortex and C indicates
the Wilson loop and $d_{r}$ is the dimension of the representation and
$\{H_{i},i=1,2,...,N-1\}$  are the generators spanning the Cartan
sub-algebra. $f_{n}$ is the
probability that any given unit is pierced by a vortex type $n$. To better understand and to compare this model with the original thin vortices, we plot $Re{\cal G}_{r}[\vec{\alpha}]$ versus $x$, for a typical $\alpha_{R}(x)$ in figure 1 (left). $\alpha_{R}(x)$ for SU(3) gauge group is defined:
\begin{equation}
\alpha_{R}(x)= \frac{2\pi}{\sqrt3} [1-\tanh(ay(x)+ \frac{b}{R})],
\label{alphar} \qquad
 y(x)= \left \{ \begin{array}{ll}
x-R ~~ \mbox{for $|R-x| \leq |x|$ } \\
-x ~~ \mbox{for $|R-x| > |x|$ } .
\end{array}
\right.
\end{equation}
$R$ is the distance between the quark and antiquark source. 
$a$ and $b$  are free parameters of the model.
The flux is plotted for $a=0.05$, $b=4$ and $R=70$.  $Re{\cal G}_{r}[\vec{\alpha}]$ changes from $1$, where there is no interaction between the Wilson loop and the vortices, to $-0.5$ where the vortex is located completely inside the loop. If we were plotting $Re{\cal G}_{r}[\vec{\alpha}]$
for the thin vortex model, we would have gotten only $1$ and $-0.5$. Thus, increasing the thickness of the vortex, implies a distribution for the vortex profile. 
We have  studied the effect of those contributions which are not exactly equal to the center elements, especially those which are close to $1$ and are not supposed to have any role in the linear part  of the potentials. This could be done by studying the effect of changing the vortex core size.
In the next section, we discuss this idea in more details and give some of the primary results.

\begin{figure}[]
\vspace{2pc}
{\includegraphics[height=.26\textheight]{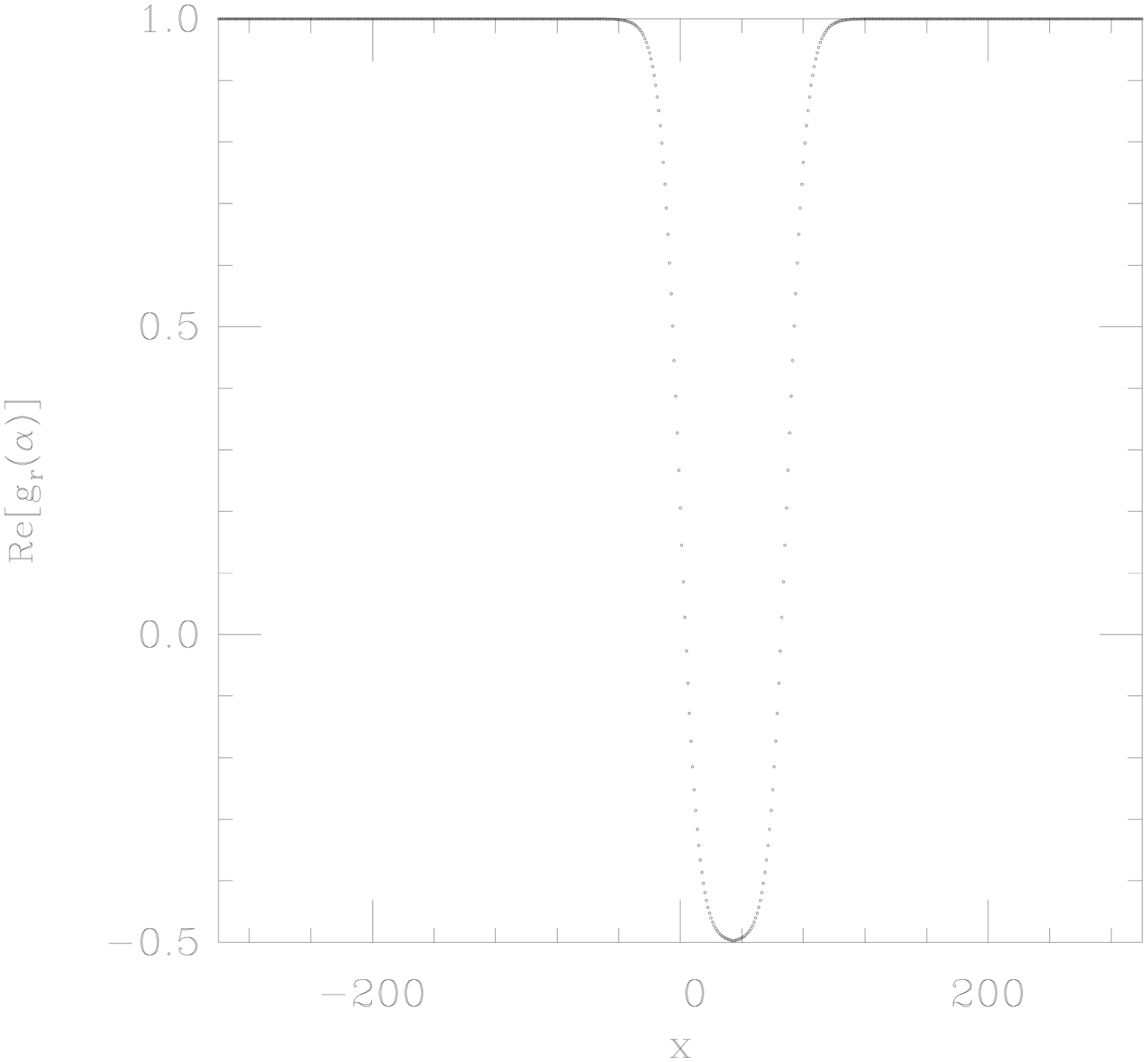}}
  \hspace{2pc}
  {\includegraphics[height=.26\textheight]{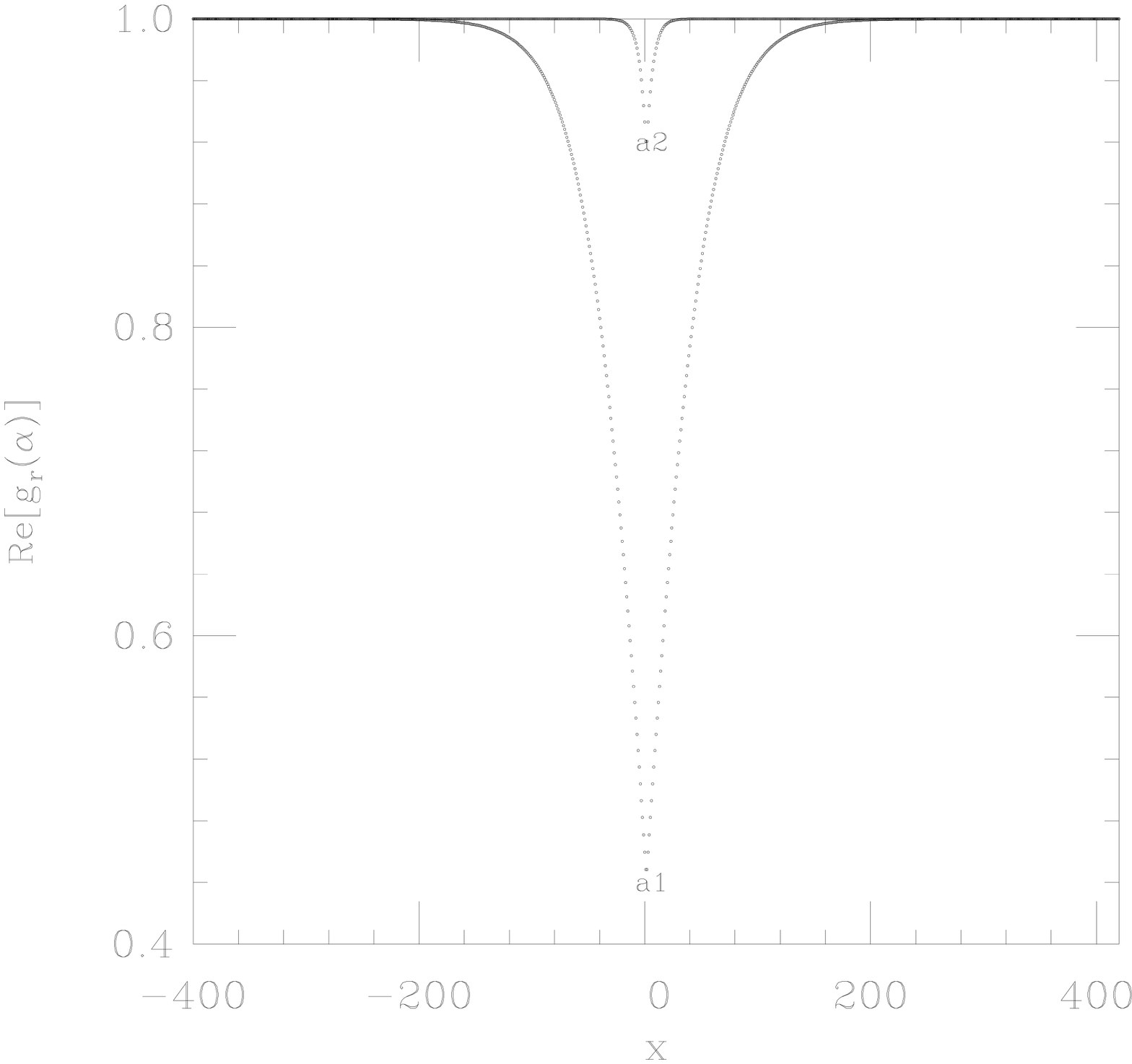}}
\vspace{-2.5pc}
\caption{On the left, $Re {\cal G}_{r}[\vec{\alpha}]$ is plotted versus $x$ for the fundamental representation of the SU(3) gauge group. $R=70$, $a=0.05$, $b=4$ and $Re {\cal G}_{r}[\vec{\alpha}]$ is equal to $1$ when no vortex links the loop and it is $-0.5$ when the vortex is completely inside the loop. On the right, we have the same plot for $R=4$  for two sets of parameters. The plot labeled by $a2$ refers to the parameters $a=0.05$, $b=4$ and the one labeled by $a1$ to  $a=0.02$, $b=1$. The distribution about the trivial center element is increased for the second case.  Compared with the plot on the left, since the vortex core is about $20$ for $a2$ and $50$ for $a1$, thus for $R=4$, the vortex never overlaps the loop completely and $Re {\cal G}_{r}[\vec{\alpha}]$ does not reach to the first non-trivial center element corresponding to $-0.5$.}
\end{figure}

\section{Fluctuating the vortex core size}

The motivation of increasing those distributions of $Re{\cal G}_{r}[\vec{\alpha}]$ which are close to $1$ is that, in equation (\ref{W}), $\exp[-\sigma(C)A]$ is obtained as a result of using the non-trivial center elements. Center elements commute with all other elements of the gauge group. The unit matrix or $I$ is another element of the group which commutes with other elements. On the other hand, $W_{0}$ in equation (\ref{W}) should give the Coulombic part. Since the non-trivial center elements are responsible for the linear part, one may suspect that the trivial center element may give the Coulombic part. In the thin vortex model, the trivial center element does not interact with the Wilson loop, replacing $z$ with $I$ in equation (\ref{WU}). But we have already thickened the vortices such that a flux distribution is used. Most of the contributions to $\exp[-\sigma(C)A]$ still comes from the non-trivial center elements, $-0.5$ in figure 1 (left), but as observed in the same figure, $Re{\cal G}_{r}[\vec{\alpha}]$ is not just $1$ or $-0.5$ but it changes between these two limits. We recall that this figure is just an example for the fundamental representation of the SU(3) gauge group. The data changes for other representations and other gauge groups. 

We have plotted $Re{\cal G}_{r}[\vec{\alpha}]$ for $R=4$ and two sets of parameters $a$ and $b$ in figure 1 (right). The set $a=0.02$, $b=1$ leads to the appearance of a Coulombic behavior for short distances. However, some concavity for the medium distance potential has been observed. To avoid this problem and increasing the size of the linear part, the vortex core size is fluctuated around $a=0.02$. We have used a Gaussian distribution but the tails of the Gaussian has been removed to avoid the core sizes which are very  different from the center value, $a=0.02$. The results are shown in figure 2 (left) where a Coulombic behavior is observed for all representations. For comparison, figure 2 (right) shows the potential for the adjoint(8) representation where the non-fluctuated core size with the old parameters are used. Table \ref{table1} shows the results of fitting the data of the model to equation:
\begin{equation}
V(R)=-(\frac {A}{R})+KR+B
\label{VR}
\end{equation}
A qualitative agreement with Casimir scaling is obtained for all the coefficients of the equation. 
\begin{figure}[]
\vspace{2pc}
{\includegraphics[height=.26\textheight]{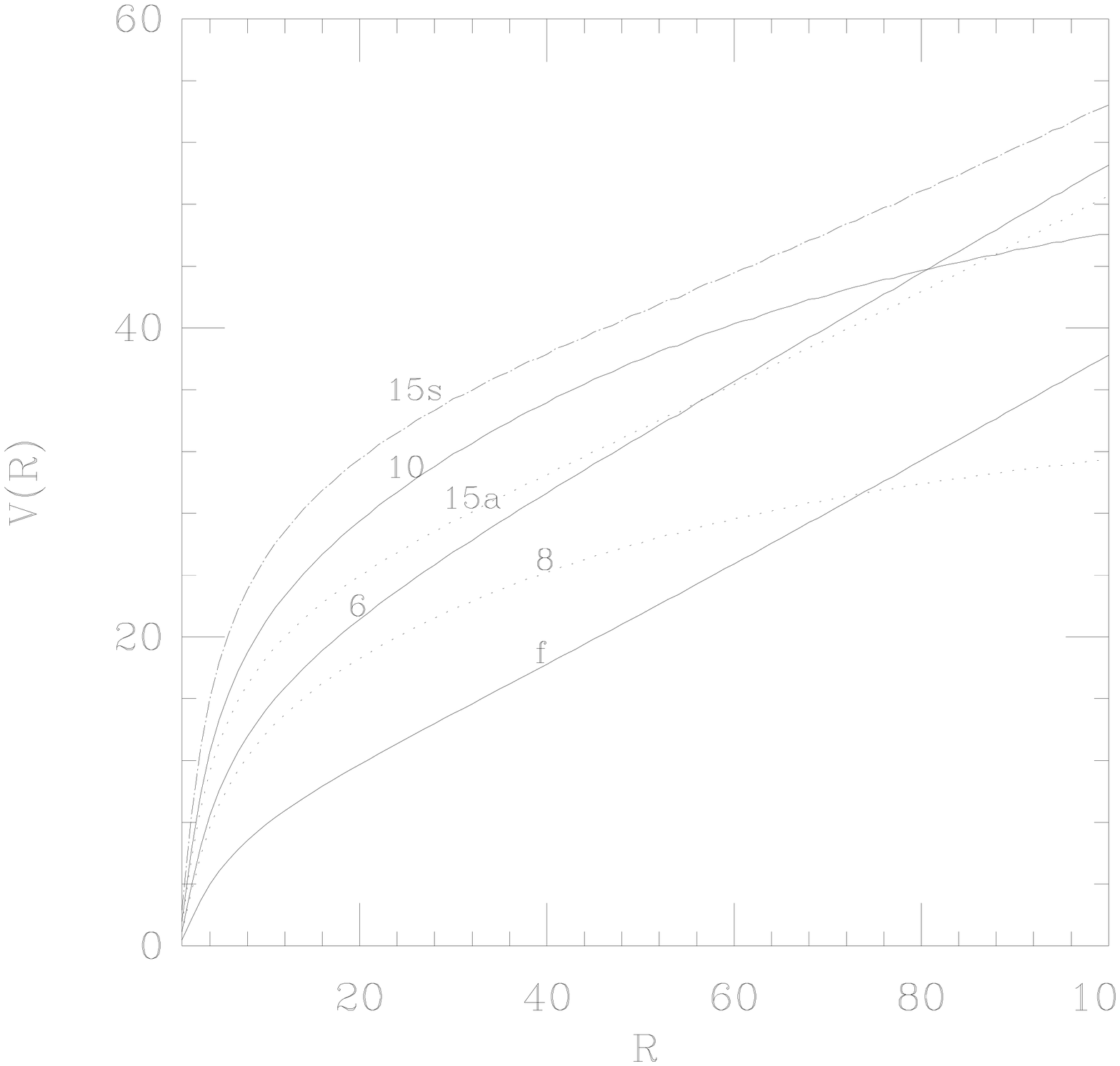}}
  \hspace{2pc}
  {\includegraphics[height=.26\textheight]{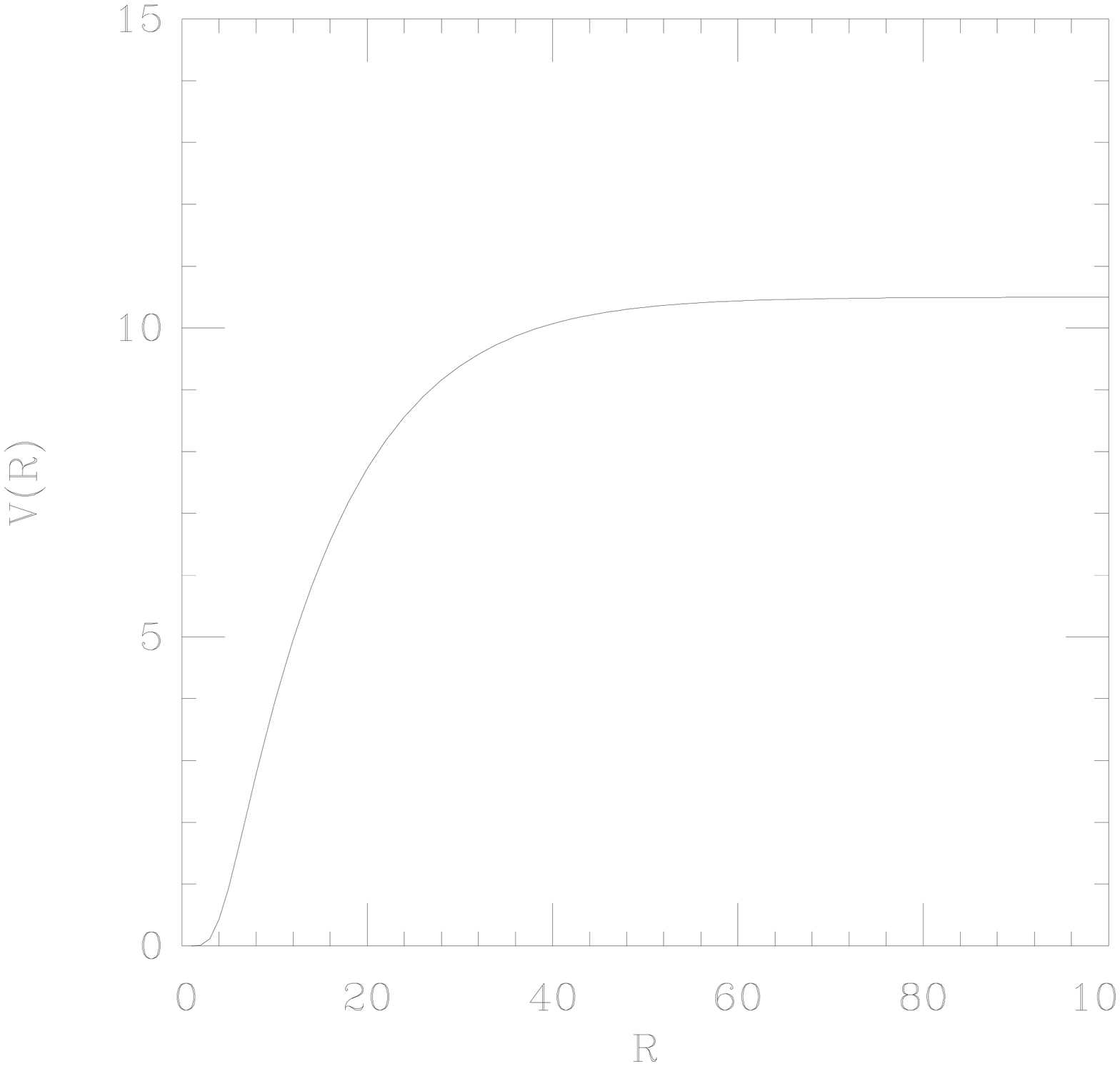}}
\vspace{-2.5pc}
\caption{On the left, potentials of some SU(3) sources are plotted versus $R$. On the right side, as an example, the potential for representation $8$ is plotted with the non-fluctuated old parameters. A Coulombic behavior, qualitatively in agreement with Casimir scaling, is seen on the left plot compared with the right one. 
}
\end{figure}
\begin{table}
\tabcolsep=2pt
\caption[]{\label{table1}The table shows the results of the fitting the data of the model to equation (\ref{VR}). $A_{r}/A_{f}$ shows the ratio of the Coulombic coefficient of representation $r$ to the fundamental representation, $k_{r}/k_{f}$ the ratio of the string tensions and $B_{r}/B_{f}$ the ratio of the constant term. $C_{r}/C_{f}$ is the Casimir ratios for the SU(3) gauge group. A qualitative agreement of the coefficients with  Casimir scaling  is
observed. The errors of the fit are shown in the parentheses.}
\begin{center}
\begin{tabular}{@{}llcccccc@{}}
\hline
Repn.&    $8$ & $6$ & $15a$&  $10$ &  $15s$
\\
\hline
$(n,m)$    & $(1,1)$ & $(2,0)$ &  $(2,1)$ & $(3,0)$ & $(4,0)$ \\
$C_{r}/C_{f}$   &  $2.25$ &  $2.5$ &  $4$ &   $4.5$ &  $7$ \\
$A_{r}/A_{f}$   &  $2.09(21)$ &   $2.31(23)$ & $2.85(27)$  &  $3.37(32)$ & $4.04(37)$\\
$k_{r}/k_{f}$   &  $1.31(8)$ &   $1.49(9)$ & $1.52(11)$  &  $1.62(10)$ &  $1.63(12)$  \\
$B_{r}/B_{f}$   &  $2.24(12)$ &   $2.45(12)$ & $3.23(15)$  &  $3.84(17)$ &  $4.93(20)$  \\
\hline
\end{tabular}
\end{center}
\end{table}
\section{Conclusion}
A Coulombic behavior for the potential between static sources is observed when the role of the trivial center element is increased in the thick center vortex model. This has been done by carefully choosing the vortex core size and fluctuating it around the chosen value. The primary results show a qualitative agreement for the Coulombic ratios with the Casimir scaling.
In general, it is very impressive that one may get the potential in all regimes with the same mechanism, thick center vortex model. We are continuing our calculations for the SU(4) gauge group.

\end{document}